\documentclass[fleqn,twoside]{article}
\usepackage{espcrc2}

\usepackage{graphicx}
\usepackage[figuresright]{rotating}

\newcommand{\AmS}{{\protect\the\textfont2
  A\kern-.1667em\lower.5ex\hbox{M}\kern-.125emS}}

\title{A new fermion Hamiltonian for lattice gauge theory}

\author{
Michael Creutz \address {Physics Department, Brookhaven
National Laboratory,\\ Upton, NY 11973, USA} \thanks{This manuscript
has been authored under contract number DE-AC02-98CH10886 with the
U.S.~Department of Energy.  Accordingly, the U.S. Government retains a
non-exclusive, royalty-free license to publish or reproduce the
published form of this contribution, or allow others to do so, for
U.S.~Government purposes.}, 
Ivan Horv\'ath \address {Dept. of Physics
and Astronomy, University of Kentucky,\\ Lexington, KY 40506,
USA}\thanks{Supported in part by DOE grant DE-FG05-84ER40154.}, and 
Herbert Neuberger \address {Dept. of Physics and
Astronomy, Rutgers University,\\ Piscataway, NJ 08855, USA}
\thanks{Supported in part by DOE grant DE-FG02-96ER40949 and by a
Guggenheim fellowship.  H. N. would like to thank L. Baulieu and the
entire group at LPTHE for their hospitality and support.}  }
       
\begin{document}

\begin{abstract}
 We formulate Hamiltonian vector-like lattice gauge theory using the
 overlap formula for the spatial fermionic part, $H_f$. We define a
 chiral charge, $Q_5$ which commutes with $H_f$, but not with the
 electric field term. There is an interesting relation between the
 chiral charge and the fermion energy with consequences for chiral
 anomalies.
\vspace{1pc}
\end{abstract}

\maketitle

\input epsf

\long \def \blockcomment #1\endcomment{}


Lattice gauge theory and chiral symmetry represent two venerable
non-perturbative approaches in particle theory.  Historically,
however, they have not easily meshed.  In particular, the old species
doubling problem forces one to insert chiral symmetry breaking
somewhere into the regulator.  The issues involved are deeply entwined
with anomalies \cite{mcreview}.  Stimulated by the domain wall fermion
idea of Kaplan \cite{kaplan} and the overlap formulation of Neuberger
and Narayanan \cite{nn}, this topic has recently seen renewed
attention and dramatic progress.

Here we explore an adaptation of the overlap idea to the Hamiltonian
formalism.  We find an interesting operator structure and a simple
intuitive picture of how anomalies work in terms of a fermion
eigenvalue flow.

To start, consider the conventional ``continuum'' Hamiltonian
for a gauge theory 
\begin{equation}\matrix{
H=H_f+H_g \hfill \cr
H_f=\overline\psi D_c \psi=\psi^\dagger \gamma_0 D_c\psi \hfill \cr
H_g=E^2+B^2 \hfill \cr
}
\end{equation}
where the fermion fields $\psi$ anticommute and
\begin{equation}
D_c=\vec\gamma\cdot(\vec\partial+ig\vec A)+m. 
\end{equation}
The operator $D_c$ consists of an antihermitean kinetic piece and the
hermitean mass term (we use hermitean gamma matrices).  The
antihermitean part anticommutes with both $\gamma_0$ and $\gamma_5$,
giving the properties
\begin{equation}
\gamma_5 D_c = D_c^\dagger\gamma_5, \hfill
\gamma_0 D_c = D_c^\dagger\gamma_0. \hfill
\end{equation}
The eigenvalues of $D_c$ lie along the line ${\rm Re}\lambda=m$ and
occur in complex conjugate pairs; i.e. if we have
$D_c \chi = \lambda \chi$, then 
$D_c \gamma_5 \chi = \lambda^* \gamma_5 \chi$.

The free theory is particularly simple in momentum space, with $D_c=
i\vec p\cdot\vec\gamma+m$.  The eigenvalues are $\lambda=\pm
i|\vec p|+m$.  This spectrum is discrete in finite volume, with
momentum quantized in units of ${2\pi\over L}$, where $L$ is the size
of our box.

Now for the lattice, ``naive'' fermions replace the momentum with a
trigonometric function $p_i \rightarrow \sin(p_i a) / a$.  This causes
the well known doubling, with extra low energy states when $p_i \sim
{\pi\over a}$.  The Wilson solution to this problem gives the mass a
momentum dependence $m\rightarrow m+{1\over a} \sum_i (1-\cos(p_i a))$
so that the doubler mass becomes of $O(1/a)$.  The free Wilson-Dirac
operator in the Hamiltonian formulation is thus
\begin{equation}
D_w= m+{1\over a}\sum_i \left(i \sin(p_i a)\gamma_i+ 1-\cos(p_i a)\right)
\end{equation}
The eigenvalues of $D_w$
lie on a superposition of ellipses, as sketched here

\bigskip
\epsfxsize .6\hsize
\centerline{\epsffile {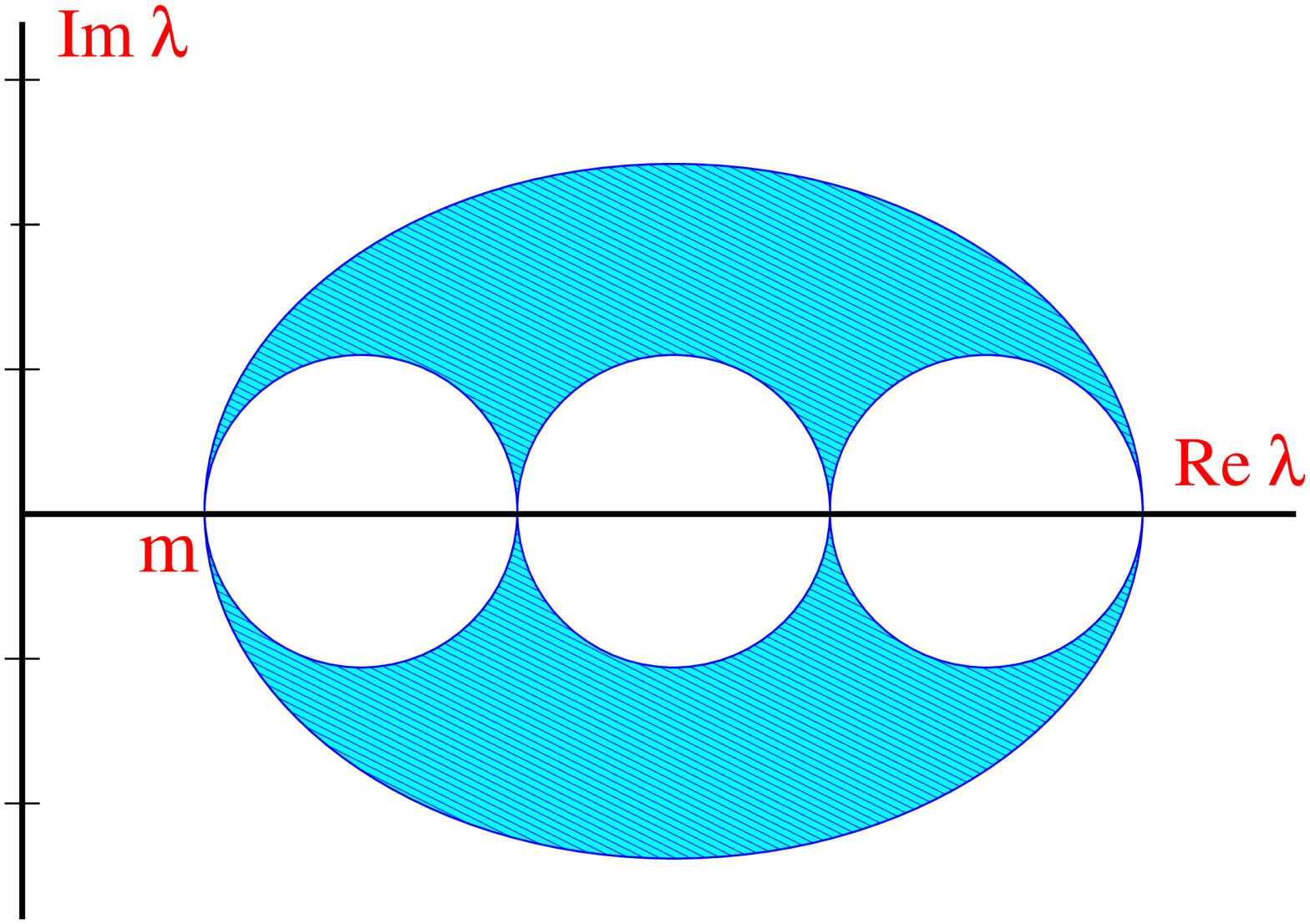}}
\bigskip

To simplify notation, from now on we work in lattice units with
$a=1$. For the continuum limit we are interested in small
eigenvalues $\lambda \sim 0$.

When gauge fields are added, several things happen.  First the fermi
eigenvalues move around, starting to fill the holes in the above
figure.  In the process the operator $D_w$ ceases to be normal, i.e
$\big[D_w,D_w^\dagger\big] \ne 0$.  The eigenvectors are no longer
orthogonal.  Also the pairing due to the $\gamma_5$ symmetry is lost.
This is directly related to the chiral symmetry violation in the
Wilson term.

To get a nicer behavior, we mimic ref. \cite{overlapoperator}
and project $D_w$ onto a unitary matrix
\begin{equation}
V= D_w (D_w^\dagger D_w)^{-1/2}
\end{equation}
Being unitary, $V^\dagger V=1$, this is a normal matrix.  From this we
construct the operator 
\begin{equation}
D=1+V
\end{equation}
and our new fermion Hamiltonian 
\begin{equation} \matrix{
H_f=\psi^\dagger h \psi \hfill\cr
h=\gamma_0 D=\gamma_0(1+V) \hfill\cr
}
\end{equation}
We recover the continuum properties
\begin{equation}
\left[D,D^\dagger\right]=0, \qquad
\gamma_5 D = D^\dagger \gamma_5
\end{equation}
and the eigenvalues again lie in complex conjugate pairs.  By
construction, the eigenvalues lie on a circle, as sketched here

\medskip
\epsfxsize  \hsize
\centerline{\epsffile {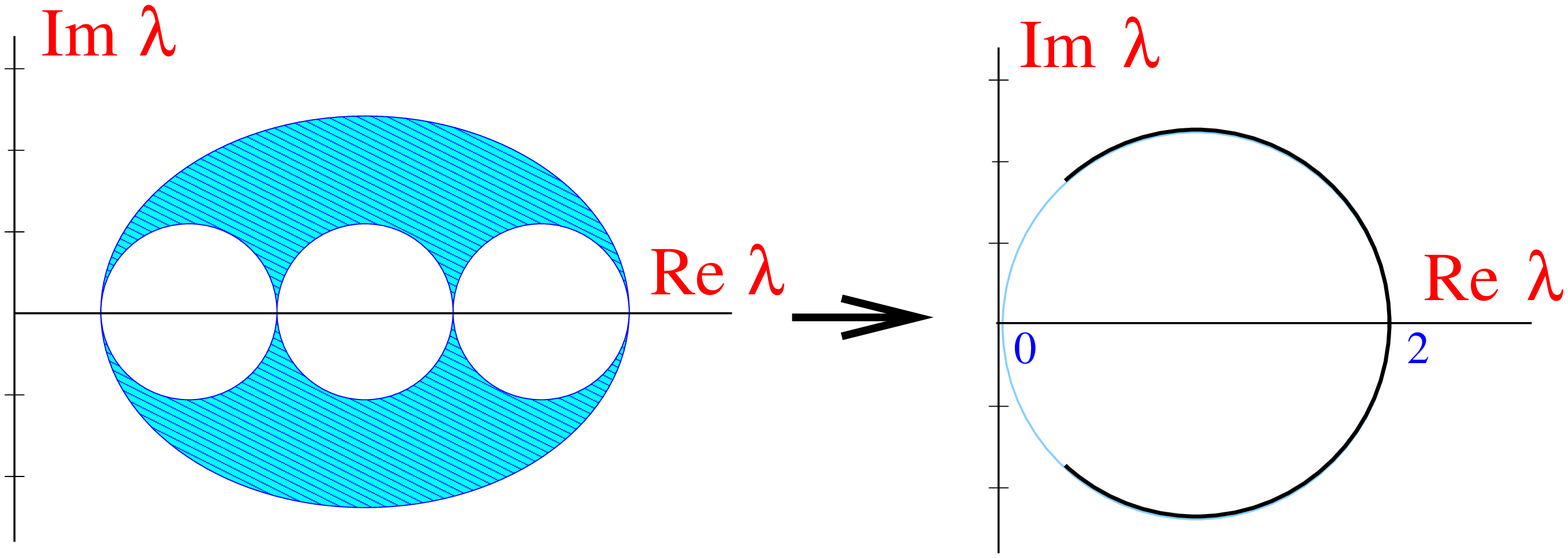}}
\medskip

As the above figure is drawn, there are no low energy states.  For
such, we must make the starting Wilson mass $m$ negative, keeping the
doubler masses positive.  Then the figure is more like this

\medskip
\epsfxsize \hsize
\centerline{\epsffile {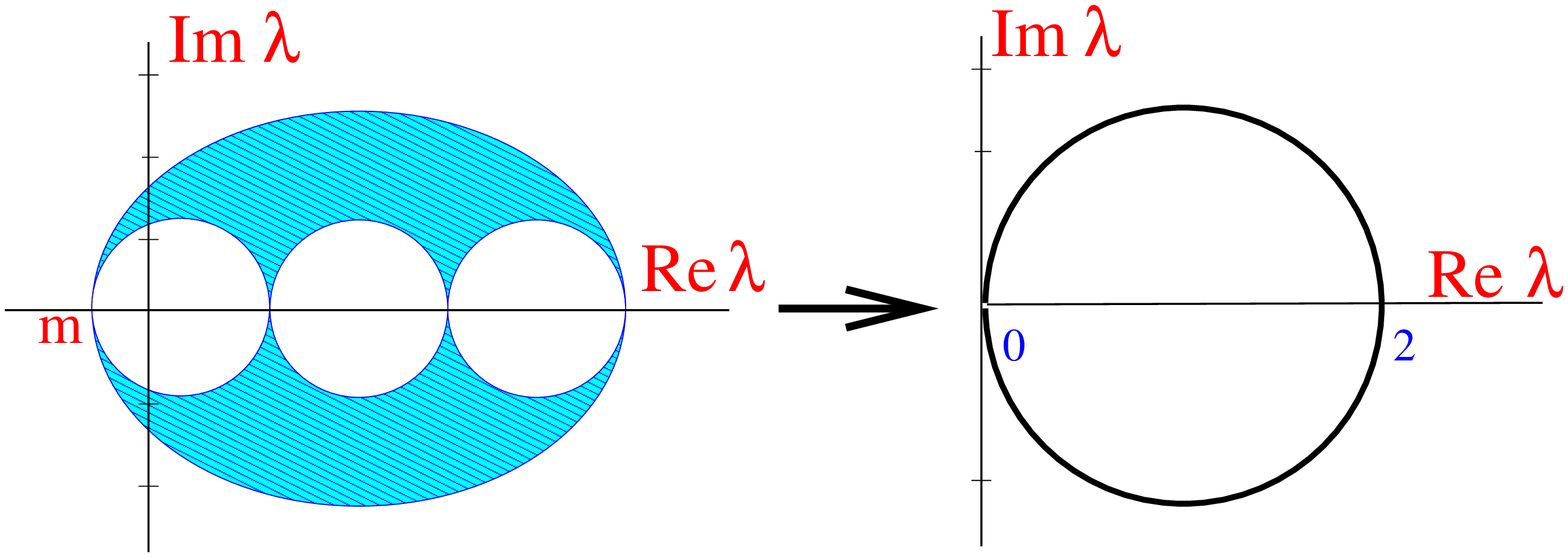}}
\medskip

As with the euclidian overlap operator, combining the unitarity
condition $V^\dagger V=1=D^\dagger D-D-D^\dagger+1$ with the
hermiticity condition $D^\dagger = \gamma_5 D \gamma_5$ gives rise to
the Ginsparg-Wilson \cite{gw} relation
\begin{equation}
\gamma_5 D + D \gamma_5-D\gamma_5 D=0
\end{equation}
However we now have another variation on this following
from $D^\dagger = \gamma_0 D \gamma_0$
\begin{equation}
\gamma_0 D + D \gamma_0-D\gamma_0 D=0
\end{equation}
By either multiplying the first of these 
two Ginsparg-Wilson relations by $\gamma_0$ or the second by
$\gamma_5$
we obtain the exact commutation relation
\begin{equation}
\big[ \gamma_5(1 - {D\over 2}),
 h\big]=0
\end{equation}
This suggests defining an axial charge
\begin{equation}\matrix{
Q_5\equiv \psi^\dagger q_5 \psi\hfill \cr
q_5\equiv \gamma_5 (1- {D\over 2}) = \gamma_5 {1-V\over 2}\hfill \cr
}
\end{equation}
This charge commutes with the fermion part of our Hamiltonian
(see also Ref.~\cite{ihht})
\begin{equation}
\big[ Q_5,H_f \big]=0.
\end{equation}

Since $D$ can be reconstructed from either $h$ or $q_5$, the matrices
$q_5$ and $h$ are closely correlated
\begin{equation}\matrix{
D=\gamma_0 h= 2-2\gamma_5 q_5\hfill \cr
\gamma_5 q_5+\gamma_0 {h\over 2}  = 1 \hfill\cr
}
\end{equation}
Squaring this gives
\begin{equation}
q_5^2+{h^2\over 4}=1.
\end{equation}
Since $q_5$ and $h$ commute they can be simultaneously diagonalized.
The above equation states that these eigenvalues lie on a circle.

Low-energy states have a well defined chirality; i.e.  $h\sim
0 \Rightarrow q_5\sim\pm 1$.  In contrast, high-energy states
all have $|q_5|<1$.  The combination $\gamma_0\gamma_5$ flips the sign
of both eigenvalues, which are thus paired on opposite sides of the
the circle.

This formulation gives a simple understanding of anomalies, analogous
to the domain-wall discussion in \cite{mcih}.  An adiabatic change of
gauge fields shifts eigenvalues continuously while the pairing is
preserved.  A topologically non-trivial shift moves an $h>0$
eigenvalue to $h<0$ while its paired state goes the other way.  In
this way a ``left'' hole and ``right'' particle (or vice versa) are
generated out of the Dirac sea, as sketched here

\medskip
\epsfxsize \hsize
\centerline{\epsffile {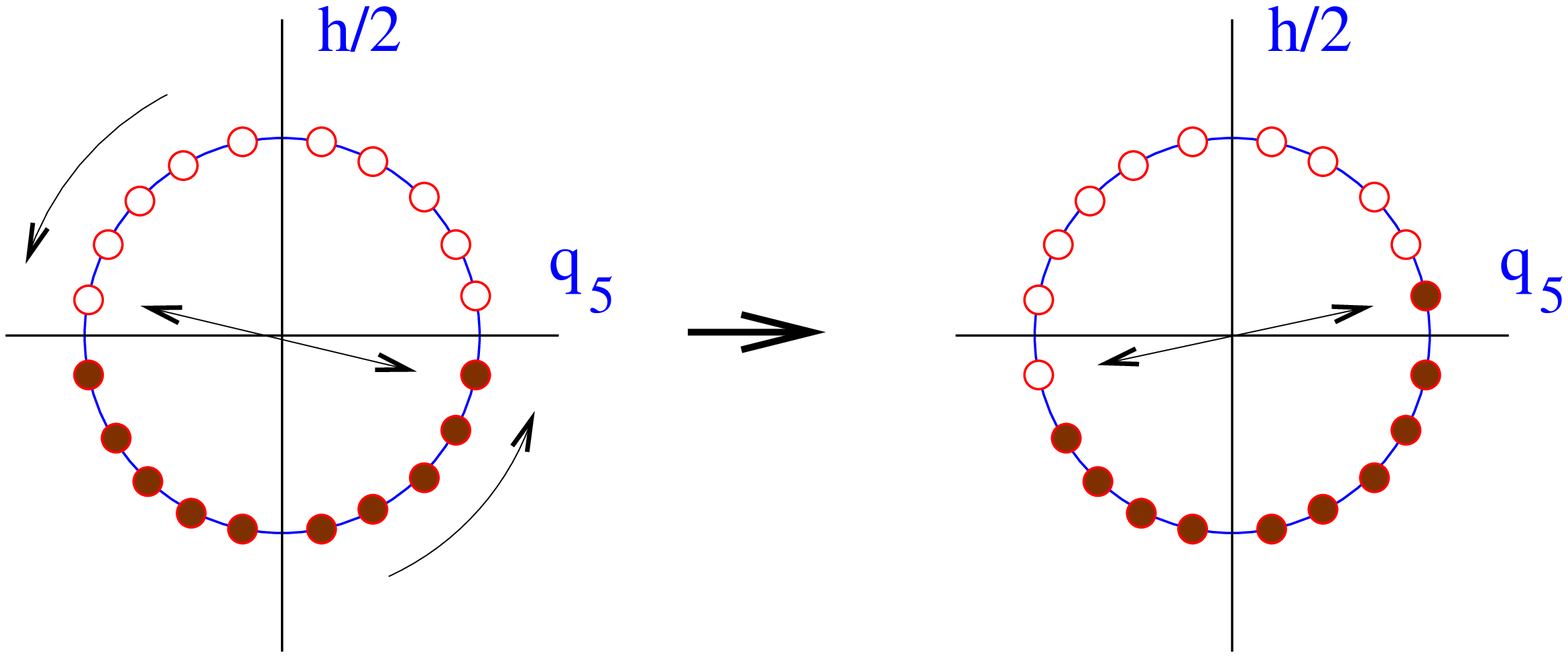}}

We have been considering the fermion states in a given background
gauge field.  For the full coupled quantum theory theory we add the
gauge field Hamiltonian
\begin{equation}
H=H_f + E^2 + B^2
\end{equation}
The operator $D$ involves link variables, which do not commute
with the electric field term, $\big[Q_5,E^2\big] \ne 0$.  This is
essential for the $U(1)$ anomaly to generate the $\eta^\prime$ mass
via mixing with gluons.

There is a close connection between these zero crossings in energy and
zero modes in Euclidean space.  Adiabatically change the gauge field
with time, ${d\over dt} h(t) = O(1/T)$ with $-{T\over 2}<t<{T\over
2}$.  Then consider an eigenvalue $h(t) \chi(t) = E_i(t) \chi(t)$
which changes in sign, $E(-{T\over 2})<0$, $E({T\over 2})>0$.  From
this construct
\begin{equation}
\phi(t)=e^{-E(t)t} \chi(t)
\end{equation}
This satisfies
\begin{equation}
D_4\phi\equiv\gamma_0\left(\partial_0+h(t)\right) \phi(t)
= 0+ O(1/T)
\end{equation}
This wave function is normalizable if the eigenvalue rises through 0.

Because of $\gamma_5$ hermiticity, $\gamma_5 D_4
\gamma_5=D_4^\dagger$, complex eigenvalues of $D_4$ are paired;
thus, unpaired zero modes are robust.  This is the lattice version of
the index theorem.

A variety of questions remain.  One involves flavored axial charges
such as $Q_5^\alpha= \psi^\dagger \tau^\alpha\gamma_5 (1-{D\over 2})
\psi$.  It appears that these also do not commute with the electric
field term of the Hamiltonian, $\big[Q_5^\alpha,E^2\big]\ne 0$.  Why
is this so, despite the euclidean overlap formulation having an exact
flavored chiral symmetry?  Another question arises at the level of
currents.  Since $Q_5$ is not ultra locally defined \cite{ivan}, what
is the natural associated $\vec J_5$?  Finally, our construction was
for vectorlike theories.  Can something similar be done for chiral
gauge theories?  In these cases anomalies change species as in the
t'Hooft \cite{thooft} baryon decay process.  For the standard model,
how do the quark and lepton ``circles'' interact to give this
phenomenon?

\end{document}